\documentstyle[epsfig]{article}

\def\be{\begin{eqnarray}}
\def\ee{\end{eqnarray}}
\def\bea{\begin{array}}
\def\eea{\end{array}}
\def\bei{\begin{itemize}}
\def\eei{\end{itemize}}

\begin{document}

\begin{flushright}
SUNY-NTG-03/03
\end{flushright}
\vspace*{2cm}
\centerline{\bf\huge  DEVELOPMENTS IN RANDOM MATRIX THEORY}
\vspace*{1cm}
\begin{center}P.J. Forrester$^1$, N.C. Snaith$^2$ and J.J.M.  Verbaarschot$^3$\\
\vspace*{1.5cm}
 $^1$Department of Mathematics and Statistics,\\
University of Melbourne, Victoria 3010, Australia\\
email: P.Forrester@ms.unimelb.edu.au\\
\vspace*{0.2cm}
$^2$School of Mathematics, University of Bristol,\\
University Walk, Clifton, Bristol BS8 1TW, UK\\
email: N.C.Snaith@bristol.ac.uk\\
\vspace*{0.2cm}
$^3$Department of Physics and Astronomy, SUNY Stony Brook, \\
Stony Brook, NY 11790, USA\\
e-mail: jacobus.verbaarschot@stonybrook.edu
\end{center}
\vspace*{2cm}
\centerline{\bf Abstract}
\vspace*{0.5cm}
In this preface to the Journal of Physics A, Special Edition
on Random Matrix Theory, we give a review of the main historical
developments of random matrix theory. A short summary of the papers that
appear in this special edition is also given.

\vfill\eject\newpage
\section{Introduction}

Random matrix theory has matured into a field with applications in
many branches of physics and mathematics. A large number of
physicists and mathematicians have been fascinated by this subject
that has turned out to be surprisingly rich and far reaching.
Paraphrasing Dyson, random matrix theory is a new kind of
statistical mechanics where  the realization of the system is not
relevant. Instead of having an ensemble of states we have an
ensemble of Hamiltonians. Ergodicity is now the equivalence of
spectral averaging and the averaging over this ensemble.

Random matrix theory has been particularly successful in three
areas: first, in describing level correlations on the scale of the
average level spacing; second, in providing the generating
function for combinatorial factors of planar diagrams; and, third,
as an exactly solvable model with intricate connections to the
theory of integrable systems. One of the reasons for the success
of random matrix theory is universality: eigenvalue correlations
on the scale of the average level spacing do not depend on the
probability distribution. This property is at the very foundation
of random matrix theory. It suggests that random matrix theory
correlations of eigenvalues should be the rule rather than the
exception. However, the most important reason for studying random
matrix theory is that its predictions do occur in nature in
systems as varied as nuclear energy levels, zeros of the Riemann
$\zeta$ function and sound waves in quartz crystals. Another of
the roles of random matrix theory is that the large $N$ limit of
its partition function is a generating function for planar
diagrams which have played an important role in quantum field
theory. For example, they are the leading contributions to QCD
with a large number of colors, and they are dual to triangulations
of a random surface and thus describe two-dimensional quantum
gravity. In addition to this, random matrix theory has attracted a
great deal of interest because of the mathematical challenges it
poses. The problems are highly nontrivial, but, with sufficient
effort, many of the questions that arise in  this field can be
answered in full.

It is hard not to be fascinated by random matrix theory. Everyone
who works in this field has experienced the amazement of obtaining
truly universal behavior by diagonalizing a large random matrix.
Nowadays, as we can see from the contributions to this special
edition, the excitement about the subject is still as much alive
as when it was first created. In the introduction of this special
edition we give a short summary of the history of random matrix
theory. The historical perspective is certainly colored by our
personal experience, and, for a somewhat different perspective, we
refer to the introduction in the review by the Heidelberg group
\cite{guhr-rev}.
We conclude with a short overview of the papers
that will appear in this special issue.

\section{History of Random Matrix Theory}

Random matrix theories have fascinated
both mathematicians and physicists since they were first introduced
in mathematical statistics by Wishart in 1928 \cite{wishart}.
After a slow start,
the subject gained prominence when Wigner \cite{wigner-50}
introduced the concept
of statistical distribution of nuclear energy levels in 1950. However,
it took until 1955 before Wigner \cite{wigner-55} introduced
ensembles of random matrices.
In this paper he also introduced the large $N$ expansion and realized
that the leading order contribution to the expectation values of moments
of the random Hamiltonian is given by  planar diagrams.
In 1956, Wigner \cite{wigner-56}
derived the Wigner surmise from the level spacing
distribution of an ensemble of $2\times 2$ matrices after
level repulsion had been predicted by Landau and Smorodinksy
\cite{landau}
and observed by
Gurevich and Pevsner \cite{gurevitch}.
The idea of invariant random matrix ensembles
was introduced in physics by Porter and Rosenzweig \cite{porter-ros}
after it had
appeared earlier in the mathematical literature. A mathematically rigorous
analysis of spacing distributions was first given by Gaudin
\cite{ga-60} and Mehta \cite{mehta-poly}.
To analyze the eigenvalue density Mehta \cite{mehta-poly1}
invented the orthogonal polynomial method.

The  mathematical foundations of
random matrix theory were established in a series
of beautiful papers by Dyson \cite{dyson-1}--\cite{dyson-5}.
He introduced the classification of random matrix ensembles according
to their invariance properties under time reversal
\cite{dyson-1,dyson-5}. As we all know, only
three different possibilities exist:  a system is not time reversal
invariant, or a system is time reversal invariant with the square of the
time reversal invariance operator either equal to 1 or $-1$.
The matrix elements
of the corresponding random matrix ensembles are complex, real and
self-dual quaternion, respectively, which from a mathematical viewpoint
exhaust the distinct real commutative normed division algebras, or in
effect number systems. The corresponding invariant
Gaussian ensembles of Hermitian random matrices
are known as the Gaussian unitary ensemble (GUE), the Gaussian
orthogonal ensemble (GOE) and the Gaussian symplectic ensemble (GSE),
in this order.

Dyson \cite{dyson-1} also formulated the underlying philosophy of
random matrix theory. In his words, ``What is here required is a 
new kind of statistical mechanics, in which we renounce exact
knowledge not of the state of the system but of the system itself.
We picture a complex nucleus as a ``black box'' in which a large
number of particles are interacting according to unknown laws. The
problem then is to define in a mathematically precise way an
ensemble of systems in which all possible laws of interaction are
equally possible''. This was made more precise by Balian \cite{balian}, who
obtained the Gaussian random matrix ensembles from minimizing the
information entropy.  

A second important result of Dyson's papers \cite{dyson-3,dyson-4}
was the relation between
random matrix theory and the theory of exactly integrable systems: the
partition function of a random matrix ensemble is equivalent to the
partition function of a log-potential
Coulomb gas in one dimension at three special temperatures, each
with solvability properties not shared for general temperatures.
Moreover the evolution of the eigenvalues of parameter dependent
extensions of the Gaussian ensembles was shown to be controlled by a
Fokker-Planck operator which also specifies the Brownian evolution of
the Coulomb gas.
These results were
further explored by Sutherland \cite{sutherland}
when he realized that the Calogero-Sutherland
quantum many body system \cite{calogero,sutherland},
for which the Hamiltonian can be constructed from $N$ independent
commuting operators and so is integrable,
is mathematically equivalent to Dyson's Brownian motion model.
The relation between random matrix theory and integrable systems is
discussed extensively in the monograph by Forrester \cite{peter}.
A review of one dimensional integrable systems that touches on many ideas
that also appear in random matrix theory is given in the book
by Korepin, Bogoliubov and Izergin \cite{korepin}. 
A third idea that appeared in Dyson's
paper \cite{dyson-1} is the application of Shannon's information entropy
to random matrix spectra.

The early developments in random matrix theory are well summarized in
the first edition of
the monograph by Mehta \cite{mehta-book-66}. This has been a very
influential book containing many mathematical details which have been
proved to be extremely useful over the years.
A second significant book is by
Porter \cite{porter-book}. It contains reprints of the important
papers on random matrix theory that were written before 1965.

About the same time as the early development of random matrix
theory in nuclear physics, the field of disordered systems was born
by the work by Anderson \cite{anderson}
on the localization of wave functions
in one dimensional disordered systems. He considered a one dimensional
lattice with a random potential at each lattice point. He found that
the eigenfunctions of this system are exponentially localized. His work
had strong impact on both experimental and theoretical solid state physics.
Another early application of random matrix theory 
is the theory of small metallic particles by Gorkov and
Eliasberg \cite{gorkov}, which
nowadays would be part of mesoscopic physics.

Random matrix theory, which was first formulated in mathematical
statistics, continued to develop in mathematics independently of
the developments in physics. Important results with regards to the
integration measure of invariant random matrix ensembles were
obtained by L.K. Hua \cite{hua-book}. His results of more than a
decade of work are summarized in his book that appeared in 1959
but which remained largely unknown. Only a small number of
mathematicians worked on integrals that appear in random matrix
theory. One very important result was obtained by Harish-Chandra
\cite{hc}, who evaluated a unitary matrix integral that is now known as the
Harish-Chandra-Itzykson-Zuber integral \cite{hc,iz}.
Zinn-Justin and Zuber \cite{zinn-justin} review this topic in the
present special issue.
Also the work
of Selberg \cite{selberg} is well-known, not in the least because
Madan Lal Mehta devoted a chapter of the second edition of his
book \cite{mehta-book-91} to this subject. Another noteworthy
contribution is the introduction of zonal polynomials by James
\cite{james}. The 1982 book of Muirhead  \cite{muirhead} ties
together matrix integrals and zonal polynomials as they are
relevant in mathematical statistics. Girko has written a number of
mathematical books (see e.g.~\cite{girko}) relating to analytic
properties of the eigenvalue distribution of large random
matrices. Voiculescu \cite{voiculescu} used random matrices as a
primary example of the concept of free non-commutative random
variables in operator algebras. However, the mathematical
literature remained largely  unnoticed by physicists until
recently.

What is more surprising is that
the theory of disordered systems and the application of
random matrix theory in nuclear physics proceeded more or less
independently until the seminal work by Efetov on the supersymmetric method
\cite{efetov-adv}
and its application \cite{efetov-small}
to the theory of small metallic particles
and to localization theory
\cite{efetov-loc} .  
This is even more remarkable since
both the papers by Anderson and Dyson were written at Princeton.

The main developments in random matrix theory  in the decade after 
the appearance of the first edition of Mehta's book
were applications to nuclear physics. In particular, the
statistical theory of $S$-matrix fluctuations received a great deal of
attention.
The first work in this direction dates back to Wigner \cite{wigner-R},
who simultaneously
studied the distribution of the widths and the spacings of nuclear resonances,
and to
Porter and Thomas \cite{porter-thomas},
who introduced the Porter-Thomas distribution for
nuclear decay widths.
Correlations of cross-sections at two different energies were considered in
\cite{torleif} and
are now known as Ericson fluctuations. 
The formulation of the theory of $S$-matrix fluctuations was
completed in the work of Agassi et al. \cite{agassi} . In this
paper the authors introduced resummation techniques which later
received much more attention in the field of impurity scattering,
introduced earlier in the book by Abrikosov, Gorkov and
Dzyaloshinski \cite{impurity-scattering}. The problem of the
distribution of poles of $S$-matrices was also the motivation of
Ginibre \cite{ginibre} for introducing what is now known as the
Ginibre ensemble with
eigenvalues          
uniformly distributed inside a disk in the complex plane.
His paper initiated the subfield of nonhermitian random matrix theory
which is reviewed by Fyodorov and Sommers \cite{fyodorov} in this issue.

In 1973 Montgomery \cite{montgomery} made a conjecture for the
asymptotic limit of the two-point correlation function of the
zeros of the Riemann $\zeta$ function on the critical line.
Together with Dyson he realized that his conjectured result is the
two-point function of the GUE.  The connection was extended to
higher correlation functions of the Riemann zeros by Hejhal
\cite{hejhal1} and Rudnick and Sarnak \cite{rudnick-sarnak},
although the full correspondence of the correlation functions with
random matrix theory has still not been proved.  A heuristic
derivation of these results using the Hardy-Littlewood conjecture
for the correlation between primes, was given by Bogomolny and
Keating in 1995 \cite{bog-kea1,bog-kea2}. Mathematically rigorous
results relating the two-point functions for the zeros of families
of finite field zeta functions and eigenvalues of random matrices
from the classical groups are the topic of the monograph by Katz
and Sarnak \cite{katz-sarnak}.

The conjectured correspondence of the statistics
of these zeros of the Riemann $\zeta$ function with the 
$n$-point correlation function of random matrix eigenvalues has
recently meant that random matrix theory has become very useful
for conjecturing quantities in number theory that were previously
unattainable by any method.  These include mean values of the
Riemann zeta function and other $L$-functions
\cite{cfkrs}--\cite{keasna00b}, the order of vanishing at special
values of $L$-functions \cite{ckrs00}, as well as discrete moments
of the derivative of the Riemann zeta function \cite{hughes00,
hughes} and the horizontal distribution of the zeros of the
derivative \cite{mezzadri}.  For more details there is a review in
this issue by Keating and Snaith \cite{nina}.

In the period 1975-1985, random matrix theory developed rapidly and became
unified with the theory of disordered systems. The first step in this direction
was made by Edwards and Anderson \cite{edwards-spin}
who, in their influential paper on spin
glasses, introduced the replica trick. This provided a natural framework
for a field theoretical formulation of the Anderson model which was introduced
a few years later by Wegner in 1979 \cite{wegner-79}.
In this formulation, symmetries and the
spontaneous breaking of symmetries led to a new paradigm in the
theory of Anderson localization  
\cite{Efetov-Larkin,McKane-Stone,Schaefer-Wegner}. It
was soon realized that the replica formulation only works well for
perturbative calculations. 
This problem was solved by the introduction of the supersymmetric
method \cite{efetov-adv}. In this method the determinants in the generating
function of the resolvent are quenched by taking
a ratio of two determinants instead of the $n\to 0$ limit
of the $n$'th power of the determinant. 
Relying on earlier work by Wegner \cite{wegner-79} using the replica trick,
Efetov showed that the partition function of a disordered system
is given by a supersymmetric nonlinear $\sigma$-model. He identified a
domain of energy differences
where the kinetic term of the non-linear $\sigma$ model can be neglected.
In this domain
the two-point correlation functions coincide with the results derived by
Dyson. The energy scale below which the partition function
is dominated by zero momentum modes is known as the Thouless
energy \cite{thouless}.

The supersymmetric method has been very fruitful. Efetov
\cite{efetov-adv} obtained new results for one-dimensional
disordered wires.  Exact results were obtained for the theory of
$S$-matrix fluctuations \cite{VWZ}. Relations between the
orthogonal and symplectic symmetry classes were derived
\cite{wegner-relations} from the supersymmetric partition
function. In the subsequent years many more new results were
derived by means of the supersymmetric method. Among others we
mention results for parametric correlations \cite{parametric}
where eigenvalue correlations for different values of an external
parameter are considered. An elaborate discussion of applications
of the supersymmetric method to disordered systems is given in the
book by Efetov \cite{efetov-book}.

Exact results for $S$-matrix fluctuations were obtained in a completely
independent way by a Mexican group \cite{pedro}.
The exact distribution function of
$S$-matrices was found starting from the three assumptions of analyticity,
ergodicity and maximizing the information entropy \cite{balian}.
Another effort in nuclear physics was the introduction of random matrix
ensembles that reflected the few-body nature of the interaction. In particular,
French and co-workers have pursued this direction of research
(see \cite{brody} for a review). In this issue Benet and Weidenmueller
\cite{benet} review recent progress in this field.

A major development was the experimental discovery of universal conductance
fluctuations by Webb and Washburn in 1986 \cite{webb}
after having been predicted theoretically by
Altshuler \cite{altshuler} and Stone and Lee
\cite{stone,stone-lee}.  
This discovery started the new field of chaotic
quantum dots. The transport properties of these quantum dots could be
described by the supersymmetric nonlinear $\sigma$-model that had been used
for the theory of $S$-matrix fluctuations in   compound nuclei. In fact,
a compound nucleus is a chaotic quantum dot
(see \cite{beenakker,yoram} and \cite{brouwer} in this issue for reviews).

A few years before the discovery of universal conductance fluctuations,
random matrix theory was applied to quantum field theory. Through
the work of 't Hooft \cite{gerard-74} we know that in the limit of
a large number of colors, the QCD partition function is dominated
by planar diagrams. This is also the case for the large $N$ limit
of random matrix theory. In \cite{brezin-79} this was exploited to
calculate the combinatorial factors that enter in the large $N_c$
limit of QCD by means of random matrix theory. A second innovative
idea that appeared in this paper is the formulation of the
calculation of the resolvent in random matrix theories as a
Riemann-Hilbert problem. This approach has received more attention
in the recent mathematical
literature \cite{deift}. 

Random matrix theory has had impact on several areas of quantum field theory:
lattice QCD, two dimensional gravity, the Euclidean Dirac spectrum and
the Seiberg-Witten \cite{seiberg-witten}
solution of two dimensional supersymmetric gauge theories. 
An important result is the Eguchi-Kawai \cite{eguchi} reduction. These
authors show that in the limit of a large number of colors, certain gluonic
correlation functions of pure Yang-Mills theory
can be reduced to an integral over 4 unitary matrices.
In two spatial dimensions this reduction results in an integral over
a single unitary matrix which can be evaluated in the large-$N$ limit.

A unitary matrix integral also occurs in the low-energy limit of
QCD. Because of the spontaneous breaking of chiral symmetry, its
low-energy degrees of freedom are the Goldstone modes which are
parameterized by a unitary matrix valued field \cite{weinberg}.
Below the Thouless energy for this system the kinetic term of the
effective Lagrangian can be neglected and the low-energy limit of
the QCD partition function is given by the unitary matrix integral
\cite{gasser-leutwyler}.
In this domain the eigenvalues of the Dirac operator 
are correlated according to a random matrix theory
with the additional involutive (chiral) symmetry of the QCD
Dirac operator \cite{SV,V}. 
The same symmetry is also found
in two-sublattice disordered systems
where hopping only occurs in between the sublattices \cite{Gade}.
The eigenvalue spectrum around zero of these chiral ensembles was
first derived in \cite{forrester}.
 An important difference between two-sublattice systems and
QCD is the topology of the random matrix (i.e. the number of exact zeros) and
the fermion determinant. In two-sublattice systems one is only interested in
quenched results at zero topology whereas in QCD the fermion
determinant and its zero modes are essential.
Also in the case of the chiral ensembles we have three different symmetry
classes depending on the reality content of the matrix elements. Most of
the work on
chiral random matrix theory and its applications to the Dirac spectrum
in QCD was done in the second half of the nineties (see
\cite{VW} for a review).

In the theory of disordered superconductors four more random matrix ensembles
can be introduced \cite{oppermann,altland-zirnbauer}, thus distinguishing
a total of 10 random matrix ensembles. It was
noticed by Dyson \cite{dyson-class}
that each of the three Wigner-Dyson random matrix theories corresponds
to a symmetric space. Zirnbauer \cite{martin-class}
showed that this observation can be
generalized to all 10 symmetry classes of
 random matrix ensembles with a one-to-one correspondence
to each of the large families in the Cartan classification of symmetric
spaces.

There have been other attempts to derive QCD from a matrix model.
Perhaps best known is the induced QCD partition function of
Kazakov and Migdal \cite{kazakov-migdal-induced} where the lattice
gauge field is coupled to an adjoint scalar field. The gauge field
can be integrated out by means of the Harish-Chandra-Itzykson-Zuber integral
resulting in a partition function for the eigenvalues of the
adjoint scalar field. This partition function can be evaluated by
saddle point methods in the large $N$ limit. More recently, it has
been shown that that the so-called prepotential of $N=2$
supersymmetric theory can be derived from the large $N$ limit of a
random matrix theory \cite {vafa-dijkgraaf}.

The partition function of 2d gravity
is a sum over random surfaces which can be described by means of a
triangulation \cite{david,kazakov-migdal}.
The sum over triangulated surfaces can be written
in terms of a
random matrix theory partition function. It has been conjectured
\cite{douglas-shenker,brezin-kazakov,gross-migdal}
that the double scaling limit
of this theory describes the continuum limit of the 2d gravity partition
function. This field brought two new ideas into random matrix theory:
universality \cite{ambjorn,bowick,ADMN}, 
i.e. that observables are independent of the probability
potential, and the connection with integrable systems. In the context of
quantum gravity it is natural to consider an arbitrary polynomial
probability potential.
Integrable hierarchies were obtained from
differential equations in the coefficients
of the probability potential \cite{douglas,dijkgraaf}.
A good review of this topic
was given by Di Francesco, Ginsparg and Zinn-Justin \cite{ginsparg}.

Earlier integrable hierarchies entered in a completely different
way. In 1980 it was found by the Kyoto school \cite{JMMS80} that
the probability of a gap free interval in the infinite GUE is a
$\tau$-function for a completely integrable system specifying the
isomonodromy deformation of a coupled system of linear
differential equations. This had the consequence that the spacing
distribution could be expressed in terms of a Painlev\'e V
transcendent. Later it was found that the distribution of the
largest eigenvalue in the GUE is given by the solution of a
Painlev\'e II equation \cite{tracy-widom}. This development found
application in the solution of a long standing mathematical
problem: specifying the limiting distribution of the longest
increasing subsequence length of a random permutation \cite{baik}.
In fact the sought distribution is the same as that for the
largest eigenvalue in the GUE. The increasing subsequence problem
can equivalently be formulated as the polynuclear growth model in
$1+1$ dimension \cite{PS01}, and similar relationships  with
random matrix fluctuations are also known for certain tiling
problems \cite{Jo02}.

The question of why random matrix theory works has been addressed from many
different points of view. It was realized early on that the detailed
properties of eigenvalue correlations do not depend on the specifics
of the probability distribution. One important reason for
random matrix theory to work is
already mentioned in a work of Dyson \cite{dyson-1},
asserting that if a system is sufficiently
complex, the state of the system is no longer important. However, it
took until the early eighties before it was realized that the key reason
is that the corresponding classical system is chaotic. Although there have
been a few earlier studies
relating random matrix theory
 correlations to classical chaos \cite{zaslavsky,berry},
it was formulated explicitly in a ground breaking paper by Bohigas,
Gianonni and Schmit \cite{bohigas}
who, based on a numerical study of the Sinai billiard \cite{sinai},
conjectured that level correlations on the scale of the average level spacing
are given by random matrix theory
if the corresponding classical system is chaotic. This
conjecture has been confirmed  for numerous systems. The
reverse was also shown numerically to be true: if the system is not completely
chaotic, the spectral correlations are not given by the Wigner-Dyson
ensembles \cite{cederbaum,seligman}.
Although a complete proof of this conjecture is still lacking, a
considerable amount of analytical
understanding has been obtained on the basis of a semiclassical analysis
\cite{berry-hd,bogomolny-keating}. These inter-relations mean that
random matrix theory plays an essential role in the study of
quantum chaos, a fact which is given prominence in the books by
Haake \cite{haake} and Stoeckmann \cite{stockmann}.

In this short historical overview we have seen that random matrix theory
has been applied to wide ranging fields. Its
scope has by far not yet been exhausted as illustrated by recent
publications that are as varied as applications
to financial correlations \cite{potters,plerou} and wireless
communication \cite{simon}.

\section{The Special Issue}

The present special edition represents a broad cross section of
the current activity in random matrix theory. We have subdivided
the forty-four contributions into eight different groups:
Applications to Number Theory, Applications to Statistical
Mechanics, Integrable Systems and Random Matrix Theory,
Integration Formulas,
 Mesoscopic Physics and
Disordered Systems, Non-Hermitian Random Matrix Theories, Quantum
Chaos,   and Special Random Matrix Ensembles.

The contributions to the section ``Applications to Number Theory''
address the connection between random matrix theory and the
Riemann zeta function and other $L$-functions.  In \cite{coram}
and \cite{hughes-rudnick} (this second in conjunction with
companion papers \cite{hughes-rudnick1,hughes-rudnick2}, published
elsewhere) this connection is examined through the comparison of
statistics of the $L$-function zeros with random matrix
eigenvalues, while \cite{hughes}, \cite{keating-jp} and
\cite{mezzadri} make use of a comparison between the
characteristic polynomial of a random matrix and the $L$-functions
themselves.
 The present
status of the application of random matrix theory to $L$-functions
is reviewed by Keating and Snaith \cite{nina}.

In section two we discuss applications of random matrix theory to
statistical mechanics. Three of the papers
\cite{hikami,connell,borodin} can be regarded as outgrowths of the
solution of the largest increasing subsequence problem for a
random permutation. Also there are papers on financial
correlations \cite{guhr}, random projections relating to phase
retrieval \cite{elser}, and on energy landscape statistics
\cite{degli}.

Already since the work of Dyson random matrix theories have been
interpreted as one-dimensional integrable systems. In particular,
the Calogero-Sutherland model could be solved by means of
techniques with their origin in random matrix theory. In this
section we have one paper on the BC-type Calogero model
\cite{shinsuke}. More recently it was found that some random
matrix theory partition functions are $\tau-$ functions of
integrable hierarchies. This is explored in
\cite{bertola,kazakov,kostov}. Other relations with integrable
structures are discussed in \cite{eynard}.

Different types of integration formulae relevant to random matrix
theory are discussed in section four. A very important class of
integrals are the Harish-Chanda-Itzykson-Zuber integrals which are
reviewed by Zinn-Justin and Zuber \cite{zinn-justin}. Other
unitary matrix integrals are discussed in \cite{wettig}. New
results for ratios of determinants are obtained in \cite{strahov}.

The fifth section is on mesoscopic physics and disordered systems.
Although this field is strongly rooted in experimental condensed
matter physics we did not receive any such contributions. However,
this section contains contains two theoretical reviews, one on
transport through a chaotic quantum dot by Polianski and Brouwer
\cite{brouwer} and one on wave function statistics by Merlin,
Evers and Mildenberger \cite{mirlin}. Three different disordered
systems are discussed, a two-sublattice system \cite{evangelou}, a
one-dimensional disordered system \cite{richert} and a microwave
cavity \cite{schafer}.

The sixth section is on Non-Hermitian Random Matrix Theories. We
can distinguish two different cases: weak non-hermiticity, which
is reviewed by Fyodorov and Sommers \cite{fyodorov}, and strong
non-hermiticity. In the first case the imaginary part of the
eigenvalues is of the order of the spacing of the real parts. In
the second case the real and imaginary parts of the eigenvalues
are of the same order of magnitude which allows for much richer
eigenvalue spectra than for Hermitian matrices (see \cite{holz},
\cite{zyczkowski}). In addition to the review by Fyodorov and
Sommers, eigenvalue correlations of different types of
non-Hermitian matrices are studied in \cite{gernot},
\cite{wiegmann} and  \cite{rider}. The level spacings of a class
of non-Hermitian matrices with real eigenvalues is studied in
\cite{AJ},   while \cite{DH} treats the statistical properties of
the zeros of a random analytic function which has close analogies
with complex eigenvalues.

During the past two decades, quantum chaos has been one of the
main anchors of random matrix theory. This special issue also has
a section on this topic. One important issue has been to show
(mostly numerically) that properties of simple chaotic systems can
be described by random matrix theory. In this section we have a
discussion of the eigenvalue spectra of the cat map
\cite{gamburd}, of scattering matrices on quantum graphs
\cite{kottos} and of action correlations and random matrix theory
\cite{smilansky}. Wave functions of chaotic quantum systems are
discussed in \cite{cerruti}. We also included in this section a
paper on the calculation of parametric correlations \cite{simons}
and a paper on random waves \cite{hejhal}.

The applications of random matrix theories are by far not exhausted by the
above mentioned sections. In the last section we have included random
matrix theories that are not included in the previous section. We have
included a review on embedded random matrix ensembles
by Benet and Weidenmueller \cite{benet},
and papers on
distance matrices \cite{bogomolny}, the Penner model \cite{deo},
the Wishart ensemble \cite{janik} and on
critical random matrix models \cite{kravtsov}.

\vspace*{1cm}

\noindent{\bf Acknowledgments}

This work was partially supported by the Australian Research
Council (P.J.F.), the Royal Society Dorothy Hodgkin Research
Fellowship scheme (N.S.), and the US DOE grant DE-FG-88ER40388 (J.J.M.V.).

\end{document}